\documentclass[ reprint,nofootinbib,amsmath,amssymb,aps]{revtex4-1}
\usepackage{graphicx}
\usepackage{dcolumn}
\usepackage[colorlinks,linkcolor=magenta,anchorcolor=cyan,citecolor=blue]{hyperref}
\usepackage{bm}
\usepackage[multiple]{footmisc}

\def\be{\begin{equation}}
\def\ee{\end{equation}}
\def\ba{\begin{eqnarray}}
\def\ea{\end{eqnarray}}

\def\nn{\nonumber}
\def\lf{\left}
\def\rt{\right}

\newcommand{\eq}[1]{(\ref{#1})}

\def\lf{\left}\def\rt{\right}\def\q{\theta} \def\w{\omega}  \def\y {\psi}   \def\p {\pi} \def\a {\alpha} \def\s {\sigma} \def\d {\delta} \def\f {\phi}  \def\h {\eta} \def\j {\varphi} \def\k {\kappa} \def\l {\lambda} \def\z {\zeta} \def\x {\xi}     \def\p {\pi} \def \inf {\infty}  
\def\Q{\Theta} \def\W{\Omega}     \def\S {\Sigma} \def\D {\Delta} \def\F {\Phi}      \def\grad{\nabla}\def\.{\cdot}
\def\math {\mathcal}
\begin{document}

\title{Examining the weak cosmic censorship conjecture by gedanken experiments for Kerr-Sen black holes}
\author{Jie Jiang$^{1,2}$}
\email{jiejiang@mail.bnu.edu.cn}
\author{Xiaoyi Liu$^{2}$}
\email{xiaoyiliu@mail.bnu.edu.cn}
\author{Ming Zhang$^{1,2}$}
\email{corresponding author: mingzhang@mail.bnu.edu.cn}
\affiliation{$^{1}$College of Physics and Communication Electronics, Jiangxi Normal University, Nanchang 330022, China}
\affiliation{$^{2}$Department of Physics, Beijing Normal University, Beijing, 100875, China}
\date{\today}

\begin{abstract}
In this paper, we investigate the weak cosmic censorship conjecture for the Kerr-Sen black holes by considering the new version of the gedanken experiments proposed recently by Sorce and Wald. After deriving the first two order perturbation inequalities  in the low energy limit of heterotic
string theory based on the Iyer-Wald formalism and applying it into the Kerr-Sen black hole, we find that the Kerr-Sen black hole cannot be overspun or overcharged by the charged matter collision after taking into account the second-order perturbation inequality, although they can be destroyed by the scene only considering the first-order perturbation inequality. Therefore, the weak cosmic censorship conjecture is preserved in the Kerr-Sen black hole at this level.
\end{abstract}
\maketitle

\section{Introduction}
A naked singularity leads to the invalidity of predictability and deterministic nature of general relativity. Therefore, the weak cosmic censorship conjecture (WCCC) asserting that the gravitational collapse of a body always ends up in a black hole rather than a naked singularity was raised \cite{RPenrose}. Several studies in the Einstein-Maxwell theory on the violation of WCC have been carried out since then \cite{Wald97}. In $1974$, Wald proposed a gedanken experiment to test this conjecture\cite{Wald94}. In this experiment, one starts with an extremal black hole and perturbs it with test particles or fields to check if it is possible to destroy the event horizon, which shows that an extremal Kerr-Newman (KN) black hole cannot be destroyed in this way. However, there are two important assumptions underlying this experiment. The black hole in consideration is initially extremal and the analysis was done only at the level of the first-order perturbation.  In $1999$, Hubeny pointed out that an ingoing particle can destroy the nearly extremal KN black hole [4–9]. And it has aroused researchers widespread attention to investigate it in various theories \cite{B1, B2, B3, B4, B5, B6,B7,B8,B9,B10,B11,B12,B13,B14,B15,B16,B17}.

Motivated by previous results, Sorce and Wald \cite{SW} have recently suggested that we should straightly consider the correction of the collision matter rather than the test particle. In this new version of gedanken experiment, they derived the first two order perturbation inequalities based on the Iyer-Wald formalism \cite{IW} and the assumption of the null energy condition of the matter fields. After taking these inequalities into consideration, they found that the WCCC can hold for the nearly extremal KN black holes under the second-order correction.

In the recent years, this setup has also been investigated in some other black holes, showing that the WCCC is well preserved for the nearly extremal black holes when the second-order perturbation inequality is considered \cite{An:2017phb,Ge:2017vun,jiejiang}. But it has not been tested in string theory yet. There exists an asymptotically flat black hole solution in the low energy limit of heterotic string theory, which was derived by Sen \cite{Sen}. And the rotating charged black hole solution is known as the Kerr-Sen black hole. In \cite{Siahaan, Gwak:2016gwj}, the author investigated the old version of the gedanken experiments in the Kerr-Sen black hole and he found that feeding a charged test particle into the Kerr-Sen black hole could lead to a violation of the extremal bound of this black hole when neglecting the self-force, self-energy, and radiative effects. However, the Kerr-Sen black hole has some similar physical properties to those in Einstein-Maxwell theory, while can also be distinguished in several other aspects. Similar to the Kerr-Newman black holes, whether the WCC can be restored in Kerr-Sen black holes when the second-order correction of the collision matter is taken into account is worth studying.

The rest of this paper is organized as follows. In the following section, we take stock of the Iyer-Wald formalism for general diffeomorphism-covariant theories and present the corresponding variation quantities. In Sec. \ref{sec3}, we focus on Kerr-Sen black holes in the four-dimensional heterotic string theory, and derive the relevant quantities in this case. In Sec. \ref{sec4}, we will present the setup for the gedanken experiment in its new version, and derive the first two order perturbation inequalities for the optimal first-order perturbation of the Kerr-Sen black holes. In Sec. \ref{sec5}, considering the second-order perturbation inequality, we examine the new version of the gedanken experiment to overcharge or overspin a near-extremal Kerr-Sen black hole, and compare the result to the one without second-order perturbation. Finally, the conclusion is presented in Sec. \ref{sec6}.
\\
\\
\\
\\
\\

\section{Iyer-Wald formalism and variational identities}\label{sec2}

Using the Noether charge formalism proposed by Iyer and Wald \cite{IW}, we would like to investigate the new gedanken experiment in the charged static Kerr-Sen black holes.

To begin with, let us consider a general diffeomorphism-covariant theory on a four-dimensional spacetime $\math{M}$. The Lagrangian can be given by a four-form $\bm{L}$ where the dynamical fields consist of a Lorentz signature metric $g_{ab}$ and other fields $\y$. Following the notations in \cite{IW}, boldface letters are used to represent differential forms and refer to $(g_{ab},\y)$ as $\f$ collectively. The variation of $\bm{L}$ can be given by
\ba\begin{aligned}\label{varL}
\d \bm{L}=\bm{E}_\f\d\f+d\bm{\Q}(\f,\d\f)\,,
\end{aligned}\ea
in which $\bm{E}_\f=0$ gives the equation of motion (EOM) of the field $\f=(g, \y)$ in this theory, and $\bm{\Q}$ is the symplectic potential three-form. The symplectic current three-form is defined by
\ba\begin{aligned}
\bm{\w}(\f,\d_1\f, \d_2\f)=\d_1\bm{\Q}(\f,\d_2\f)-\d_2\bm{\Q}(\f,\d_1\f)\,.
\end{aligned}\ea
After replacing $\delta$ by the infinitesimal difeomorphism  $\mathcal{L}_\zeta$ related the vector field $\zeta^a$, one can define the Noether current three-form $\bm{J}_\z$ associated with $\z^a$,
\ba\label{defJ}
\bm{J}_\z=\bm{\Q}(\f, \math{L}_\z\f)-\z\.\bm{L}\,.
\ea
Straightforwardly, we can get
\ba
d\bm{J}_\z=-\bm{E}_\f\math{L}_\z \f\,,
\ea
which implies that $J_\zeta$ is closed if we assume that the background fields satisfy the EOM. From [34], the Noether current can be expressed as
\ba\label{JQ}
\bm{J}_\z=\bm{C}_\z+d\bm{Q}_\z\,,
\ea
where $\bm{Q}_\z$ is the Noether charge and $\bm{C}_\z=\z^a\bm{C}_a$ are the theory's constraints, i.e., $\bm{C}_a=0$ if the field satisfies the EOM.

Assuming that $\zeta^a$ is a Killing vector of the background geometry and keeping $\zeta^a$ fixed by virtue of  diffeomorphism- covariance, the first two variational identities can be further obtained
\ba\begin{aligned}\label{var1}
d[\d\bm{Q}_\z-\z\.\bm{\Q}(\f,\d\f)]=\bm{\w}\lf(\f,\d\f,\math{L}_\z\f\rt)-\z\.\bm{E}_\f\d\f-\d \bm{C}_\z\,,
\end{aligned}\nn\\\ea
and
\ba\begin{aligned}\label{var2}
&d[\d^2\bm{Q}_\z-\z\.\d\bm{\Q}(\f,\d\f)]\\
&=\bm{\w}\lf(\f,\d\f,\math{L}_\z\d\f\rt)-\z\.\d\bm{E}\d\f-\d^2 \bm{C}_\z\,.
\end{aligned}\ea
Below we shall consider the Kerr-Sen spacetimes. Without loss of generality, here we consider a stationary black hole with the horizon Killing field
\ba
\x^a=t^a+\W_\math{H}\j^a\,,
\ea
where $t^a$, $\j^a$ , and $\W_\math{H}$ are the timelike Killing field, the axial Killing field and the angular velocity of the horizon, respectively. The ADM mass and angular momentum of this black hole are given by
\ba\begin{aligned}\label{dMJ}
\d M&=\int_\inf \lf[\d\bm{Q}_t-t\.\bm{\Q}(\f,\d\f)\rt]\,,\\
\d J&=\int_\inf \d\bm{Q}_\j\,.
\end{aligned}\ea
For the hypersurface $\Sigma$ connected the horizon and spacial infinity, using the Stokes theorem, we can further obtain
\ba\begin{aligned}\label{var12}
\d M-\W_\math{H}\d J&=\int_{B}\lf[\d\bm{Q}_\x-\x\.\bm{Q}(\f,\d\f)\rt]-\int_\S \d \bm{C}_\x\,,\\
\d^2 M-\W_\math{H}\d^2 J&=\int_{B}\lf[\d^2\bm{Q}_\x-\x\.\d \bm{Q}(\f,\d\f)\rt]\\
&-\int_\S\x\.\d \bm{E}\d\f-\int_\S \d \bm{C}_\x+\math{E}_\S(\f,\d\f)\,,
\end{aligned}\ea
where $B$ is the cross section on the horizon and we have denoted
\ba
\math{E}_\S(\f,\d\f)=\int_\S\bm{\w}(\f,\d\f,\math{L}_\x\d\f)\,.
\ea

\section{effective action of the heterotic string theory and Kerr-Sen black hole solution}\label{sec3}

In this section, we consider the four-dimensional Kerr-Sen black hole, which is described by the four-dimensional effective action of the low energy limit of heterotic string theory,
\ba\begin{aligned}
S&={16\p}\int d^4x\sqrt{-\math{G}} e^{-\y}\lf(\math{R}-\frac{1}{8}\math{F}^2\right.\\
&\left.-\frac{1}{12}\math{H}^2+\math{G}^{ab}\grad_a\y \grad_b \y \rt)\,,
\end{aligned}\ea
where we denote
\ba\begin{aligned}
\math{F}^2&=\math{G}^{ac}\math{G}^{bd}F_{ab}F_{cd}\,,\\
\math{H}^2&=\math{G}^{ac}\math{G}^{bd}\math{G}^{ce}H_{abc}H_{cde}\,.
\end{aligned}\ea
Here $\math{G}_{ab}$ stands for the metric in string frame which is related to the Einstein metric by $g_{ab}=e^{-\y}\math{G}_{ab}$, $\math{R}$ is the Ricci scalar of $\math{G}_{ab}$, $\y$ is dilaton field, $\bm{F}=d\bm{A}$ is the Maxwell field, and $\bm{H}=d\bm{B}-\frac{1}{4}\bm{A}\wedge d\bm{A}$ with an antisymmetric tensor gauge field $\bm{B}$. It is not hard to see that this system is invariant under the $U(1)$ gauge transformation $\bm{A}\to \bm{A}+d\a\,, \bm{B}\to\bm{B}+\frac{\a}{4} \bm{F}$ with an arbitrary scalar field $\a$. For later convenience, we transform the system into the Einstein frame. Then, the Lagrangian can be expressed as
\ba\begin{aligned}
\bm{L}=\frac{\bm{\epsilon}}{16\p}\lf(R-\frac{1}{2}\grad^a\y \grad_a \y-\frac{e^{-\y}}{8}F^2-\frac{e^{-2\y}}{12}H^2 \rt)\,,
\end{aligned}\ea
where $\bm{\epsilon}$ and $R$ are the volume element and Ricci scalar of the Einstein metric $g_{ab}$. We denote $F^2=F_{ab}F^{ab}$ and $H^2=H_{abc}H^{abc}$ in which the indexes are raised by the inverse Einstein metric $g^{ab}$. After the field redefinition (together with a rescaling $\y\to 2\y$, $\bm{A}\to 2\sqrt{2}\bm{A}$), this action will reduce to the action of the Einstein-Maxwell-dilaton gravity, except for the $H^2$ term. Therefore, in this case, the electric potential of this black hole should be defined by
\ba\label{FH}
\F_\math{H}=\frac{1}{2\sqrt{2}} \left.A_a\x^a\right|_\math{H}\,,
\ea
where $\x^a$ is the Killing vector of the Killing horizon.

In this paper, we only consider the collision matter with an electromagnetic matter source; i.e., the sources of the dilaton field $\y$ and the gauge field $\bm{B}$ always vanish, that is to say, $E_\text{B}^{ab}=E_\y =0$ even though the perturbation is taken into account. Then, the equations of motion can be written as
\ba\begin{aligned}\label{eom1}
&R_{ab}-\frac{1}{2}R g_{ab}=8\p\lf(T_{ab}^\text{A}+T_{ab}^\text{B}+T_{ab}^\text{DIL}+T_{ab}\rt)\,,\\
&\grad_a \tilde{F}^{ac}+\frac{1}{2}\lf(\tilde{H}^{abc}F_{ab}-A_a\grad_b \tilde{H}^{abc}\rt)=8\sqrt{2}\p j^c\,,\\
&E_\text{B}^{bc}=\frac{1}{32\p}\grad_a\tilde{H}^{abc}=0\,,\\
&E_\y=\frac{1}{16\p}\lf(\grad^2\y+\frac{1}{8}e^{-\y}F^2+\frac{1}{6}e^{-2\y}H^2\rt)=0\,,
\end{aligned}\ea
with
\ba\begin{aligned}\label{TTT}
T_{ab}^\text{A}&=\frac{e^{-\y}}{32\p}\lf(F_{ac}F_b{}^c-\frac{1}{4}g_{ab}F^2\rt)\,,\\
T_{ab}^\text{B}&=\frac{e^{-2\y}}{32\p}\lf(H_{acd}H_b{}^{cd}-\frac{1}{6}g_{ab}H^2\rt)\,,\\
T_{ab}^\text{DIL}&=\frac{1}{16\p}\lf(\grad_a\y\grad_b\y-\frac{1}{2}g_{ab}\grad_c\y\grad^c\y\rt)\,,
\end{aligned}\ea
where $T_{ab}$ and $j^a$ are the stress-energy tensor and electromagnetic charge current of the additional matter source. The part associated with the equation of motion in the Lagrangian variation \eq{varL} is given by
\ba\begin{aligned}\label{EC}
\bm{E}_\f(\f)\d\f&=-\bm{\epsilon}\lf(\frac{1}{2}T^{ab}\d g_{ab}+E_\text{B}^{ab}\d B_{ab}\right.\\
&\left.+\frac{1}{2\sqrt{2}}j^a\d A_a+E_\y\d \y\rt)\,.
\end{aligned}\ea
According to this Lagrangian, the symplectic potential can be decomposed into four parts
\ba\begin{aligned}
\bm{\Q}(\f,\d\f)&=\bm{\Q}^\text{GR}(\f,\d\f)+\bm{\Q}^{\text{A}}(\f,\d\f)\\
&+\bm{\Q}^{\text{B}}(\f,\d\f)+\bm{\Q}^\text{DIL}(\f,\d\f)\,,
\end{aligned}\ea
with
\ba\begin{aligned}\label{qqq}
\bm{\Q}_{abc}^\text{GR}(\f,\d\f)&=\frac{1}{16\p}\epsilon_{dabc}g^{de}g^{fg}\lf(\grad_g \d g_{ef}-\grad_e\d g_{fg}\rt)\,,\\
\bm{\Q}_{abc}^\text{A}(\f,\d\f)&=-\frac{1}{32\p}\epsilon_{dabc}\tilde{F}^{de}\d A_e\,,\\
\bm{\Q}_{abc}^\text{B}(\f,\d\f)&=\frac{1}{64\p}\epsilon_{dabc}\lf(\tilde{H}^{def}A_f \d A_e-2\tilde{H}^{def}\d B_{ef}\rt)\,,\\
\bm{\Q}_{abc}^\text{DIL}(\f,\d\f)&=-\frac{1}{16\p}\epsilon_{dabc}(\grad^d\y) \d \y\,.
\end{aligned}\ea
Here we have defined
\ba\begin{aligned}
\tilde{\bm{F}}=e^{-\y}\bm{F}\,,\ \ \tilde{\bm{H}}=e^{-2\y} \bm{H}\,.
\end{aligned}\ea
Also, we can calculate out the Noether current as
\ba
\bm J_{abc}=\bm J^{\text{GR}}_{abc}+\bm J^{\text{GAG}}_{abc}+\bm J^{\text{DIL}}_{abc},
\ea
where
\ba\begin{aligned}
&\bm J^{\text{GR}}_{abc}=\frac{1}{8\pi} \epsilon_{dabc}\grad_f (\grad^{[f}\xi^{e]})+\epsilon_{eabc}G_f^e\xi^f\,,\\
&\bm J^{\text{GAG}}_{abc}=\frac{1}{2\sqrt{2}}\epsilon_{dabc}j^d A_f\xi^f +2\epsilon_{dabc}E_B^{df}B_{gf}\xi^g\\
&\ \ -(T^\text{GAG})^{d}{}_e\xi^e\epsilon_{dabc}-\frac{1}{32\p}\epsilon_{dabc}\grad_e(F^{de}A_f\xi^f\\
&\ \ +\frac{1}{2}H^{def}A_f A_g \xi^g+2H^{def}\xi^g B_{gf})\,,\\
&\bm J^{\text{DIL}}_{abc}
=-(T^{\text{DIL}})^d{}_e\xi^e\epsilon_{dabc}\,,
\end{aligned}\ea
in which we denote
\ba
T^\text{GAG}_{ab}=T^\text{A}_{ab}+T^\text{B}_{ab}\,.
\ea
According to \eq{JQ}, the above results imply that the Noether charge and constraint are given by
\ba\begin{aligned}
\bm{Q}_\x&=\bm{Q}_\x^\text{GR}+\bm{Q}_\x^\text{GAG}\,,\\
\bm{C}_{abcd}&=\epsilon_{ebcd}\lf(T_a{}^e+\frac{1}{2\sqrt{2}}A_aj^e+2E_B^{df}B_{gf}\xi^g\rt)\,,
\end{aligned}\ea
where
\ba\begin{aligned}
\lf(\bm{Q}_\x^\text{GR}\rt)_{ab}&=-\frac{1}{16\p}\epsilon_{deab}\grad^d\x^e\,,\\
\lf(\bm{Q}_\x^\text{GAG}\rt)_{ab}&=-\frac{1}{16\p}\epsilon_{deab}(F^{de}A_f\xi^f\\
+\frac{1}{2}&H^{def}A_fA_g\xi^g+2H^{def}\xi^gB_{gf})\,.
\end{aligned}\ea

The flux of the stress-energy tensors for the matter field must be vanishing under a stationary spacetime background. From \eq{TTT}, it implies that $F^{ab}$ must satisfy
\ba
\x^aF_{ab}\propto \x_b\,,
\ea
the antisymmetric tensor $\bm{H}$ must be purely tangential to the horizon, and the dilaton field $\y$ satisfies $\math{L}_\x\y=0$, with the tangent tensor $w^{ab}$ of the horizon and normal vector $k^a$ of the horizon\cite{SW}.

From \eq{qqq}, the symplectic current for the low energy limit of heterotic string theory can be written as
\ba\label{Fform}
\bm{\w}_{abc}(\f, \d_1\f,\d_2\f)=\bm{\w}_{abc}^\text{GR}+\bm{\w}_{abc}^\text{A}+\bm{\w}_{abc}^\text{B}+\bm{\w}_{abc}^\text{DIL}\,,
\ea
where
\ba\begin{aligned}\label{3w}
&\bm{\w}_{abc}^\text{GR}=\frac{1}{16\p}\epsilon_{dabc}w^d\,,\\
&\bm{\w}_{abc}^\text{A}=\frac{1}{32\p}\lf[\d_2\lf(\epsilon_{dabc}\tilde{F}^{de}\rt)\d_1 A_e -\d_1\lf(\epsilon_{dabc}\tilde{F}^{de}\rt)\d_2 A_e \rt]\,,\\
&\bm{\w}_{abc}^\text{B}=-\frac{1}{64\p}\lf[\d_2\lf(\epsilon_{dabc}\tilde{H}^{def}A_f\rt)\d_1 A_e \right.\\ &\left.-\d_1\lf(\epsilon_{dabc}\tilde{H}^{def}A_f\rt)\d_2 A_e-2\d_2\lf(\epsilon_{dabc}\tilde{H}^{def}\rt)\d_1 B_{ef}\right.\\
&\left.+2\d_1\lf(\epsilon_{dabc}\tilde{H}^{def}\rt)\d_2 B_{ef}\rt]\,,\\
&\bm{\w}_{abc}^\text{DIL}=\frac{1}{16\p}\lf[\d_2\lf(\epsilon_{dabc}\grad^d\y\rt)\d_1 \y-\d_1\lf(\epsilon_{dabc}\grad^d\y\rt)\d_2 \y \rt]\,,
\end{aligned}\ea
in which we denote
\ba\begin{aligned}
w^a=P^{abcdef}\lf(\d_2g_{bc}\grad_d\d g_{ef}-\d_1 g_{bc}\grad_d\d_2g_{ef}\rt)\,,
\end{aligned}\ea
with
\ba\begin{aligned}
P^{abcdef}&=g^{ae}g^{fb}g^{cd}-\frac{1}{2}g^{ad}g^{be}g^{fc}-\frac{1}{2}g^{ab}g^{cd}g^{ef}\\
&-\frac{1}{2}g^{bc}g^{ae}g^{fd}+\frac{1}{2}g^{bc}g^{ad}g^{ef}\,.
\end{aligned}\ea

According to the equations of motion \eq{eom1}, we can further obtain
\ba\label{Qej}
d\bm{Q_e}=\star \bm{j}\,,
\ea
where
\ba
(\bm{Q_e})_{ab}=-\frac{1}{16\sqrt{2}\p}\epsilon_{cdab}\lf(\tilde{F}^{cd}-\tilde{H}^{cde}A_e\rt)
\ea
is the Noether charge associated with the electric charge of the spacetime. The electric charge of this black hole can be defined by
\ba\begin{aligned}\label{Qe}
Q=\int_\inf \bm{Q}_e\,.
\end{aligned}\ea

We next focus on the Kerr-Sen black hole solution. In the Einstein metric, the line element of this spacetime can be read off \cite{Sen}
\ba\begin{aligned}\label{ds2}
ds^2&=-\lf(1-\frac{2M r}{\S}\rt)dt^2+\S\lf(\frac{dr^2}{\D_\text{KS}}+d\q^2\rt)\\
&+\lf(\S+a^2\sin^2\q+\frac{2M r a^2 \sin^2\q}{\S}\rt)\sin^2\q d\f^2\\
&-\frac{4M r a}{\S}\sin^2\q dt d\f\,.
\end{aligned}\ea
The electromagnetic field $\bm{A}$, antisymmetric tensor $\bm{B}$ and the dilaton field can be expressed as
\ba\begin{aligned}\label{ABF}
&\bm{A}=\frac{2\sqrt{2}Q r}{\S}\lf(dt-a\sin^2\q d\f\rt)\,,\\
\bm{B}=&\frac{2b r a \sin^2\q}{\S} dt\wedge d\f\,,\y=-\ln \frac{\S}{r^2+a^2\cos^2\q}\,,
\end{aligned}\ea
where the metric functions are given by
\ba\begin{aligned}
&\D_\text{KS}=r(r+2b)-2 M r+a^2\,,\\
&\S=r(r+2b)+a^2\cos^2\q\,,
\end{aligned}\ea
with the parameters being
\ba\begin{aligned}
b=\frac{Q^2}{2M}\,,\ \ \ \ a=\frac{J}{M}\,.
\end{aligned}\ea
According to the expressions \eq{dMJ} and \eq{Qe}, we can further find that the parameters $M,\  Q$, and $J$ are associated with the physical mass, the charge and the angular momentum of this spacetime.  From the above expression and the line element \eq{ds2}, for the nonextremal black hole, there are two horizons which are determined by $\D=0$ and can be given by
\ba
r_\pm=M-\frac{Q^2}{2M}\pm\sqrt{\lf(M-\frac{Q^2}{2M}\rt)^2-\lf(\frac{J}{M}\rt)^2}\,.
\ea
The corresponding surface gravity, area, angular velocity, and electric potential of the horizon can be shown as
\ba\begin{aligned}\label{kAWF}
\k&=\frac{r_+-r_-}{4 M r_+}\,, A_\math{H}=4\p \lf[\lf(\frac{J}{M}\rt)^2+\frac{Q^2r_+}{M}+r_+^2\rt]\,,\\
\W_\math{H}&=\frac{J}{2M^2 r_+}\,,\ \ \F_\math{H}=\frac{Q}{2M}\,.
\end{aligned}\ea
Kerr-Sen black holes become extremal when
\ba
(2M^2-Q^2)^2-4J^2=0\,,
\ea
where the inner and outer horizons coincide with each other. If $(2M^2-Q^2)^2-4J^2<0$, this metric describes a naked singularity.

\section{Perturbation inequalities of gedanken experiments}\label{sec4}
In what follows, we consider the new version of the gedanken experiment introduced by Source and Wald in \cite{SW}. The situation we plan to study is what happens to the Kerr-Sen black holes when they are perturbed by a one-parameter family charged matter source which goes into the black hole through a finite portion of the future horizon. Then, the corresponding equations of motion can be expressed as
\ba\begin{aligned}\label{eoml}
&G_{ab}(\l)=8\p\lf[T_{ab}^\text{A}+T_{ab}^\text{B}(\l)+T_{ab}^\text{DIL}(\l)+T_{ab}(\l)\rt]\,,\\
&\grad_a^{(\l)} \tilde{F}^{ac}(\l)+\frac{1}{2}\tilde{H}^{abc}(\l)F_{ab}(\l)=8\sqrt{2}\p j^c(\l)\,,\\
&E_\text{B}^{bc}(\l)=0\,,\ \ \ \ E_\y(\l)=0\,,
\end{aligned}\ea
with $T^{ab}(0)=0$ and $j^a(0)=0$ for the background fields. Here we also assume that the collision matter does not contain the sources of the dilaton field $\y$ and the antisymmetric tensor gauge field $\bm{B}$. With a similar consideration to \cite{SW}, here we also assume the Kerr-Sen black hole is linearly stable to perturbations. That is to say, any source-free solution to the linearized equation of motion \eq{ds2} will approach a perturbation towards another Kerr-Sen black hole at sufficiently late times. According to the above setup, it is not difficult to see that $T^{ab}(\l)$ and $j^a(\l)$ are vanishing except in a compact region of the future horizon. Therefore, a hypersurface $\S=\math{H}\cup \S_1$ can be chosen, starting from the  unperturbed horizon's bifurcate surface $B$,  to the horizon through the portion $\math{H}$ until the very late cross section $B_1$ with vanishing matter sources, and getting spacelike as $\S_1$ approaches spatial infinity, where the dynamical fields can be expressed by \eq{ds2} and \eq{ABF}.

For simplification, we will denote $\h=\h(0)$ to the background quantity, and the first-order and second-order variations of the quantity can be expressed as
\ba\begin{aligned}
\d \h=\left.\frac{d \h}{d\l}\right|_{\l=0}\,,\ \ \ \ \d^2 \h=\left.\frac{d^2 \h}{d\l^2}\right|_{\l=0}\,
\end{aligned}\ea
in the above setup. With these preparations, we can get the first-order inequality that the perturbation at $\l=0$ obeys. Considering that the perturbation vanishes on $B$, the first equation of Eq. \eq{var12} can further be written as
\ba\begin{aligned}
\d M-\W_\math{H} \d J&=-\int_\S \d \bm{C}_\x\\
&=-\int_\math{H}\epsilon_{ebcd}\lf(\d T_a{}^e+\frac{1}{2\sqrt{2}}A_a\d j^e\rt)\x^a\,,
\end{aligned}\ea
where $T^{ab}=j^a=0$ has been used. On $\math{H}$, the constant $\x^a A_a$ can be pulled out from the integral kernel. $\d Q_\text{flux}=\int_\math{H}\d (\epsilon_{ebcd}j^e)$ is nothing but the total charge through the horizon. As all charge vanishes through the horizon in $\math{H}$, together with Eq. \eq{Qej}, we can know that $\d Q_\text{flux}=\d Q$. Hence, the first-order perturbation equation becomes
\ba\label{var1eq1}\begin{aligned}
\d M-\W_\math{H} \d J-\F_\math{H}\d Q&=-\int_\math{H}\epsilon_{ebcd}\d T_a{}^e\x^a\\
&=\int_\math{H}\bm{\tilde{\epsilon}} \d T_{ab}k^a \x^b\,,
\end{aligned}\ea
with $\bm{\tilde{\epsilon}}$ the volume element on the horizon, defined by $\epsilon_{ebcd}=-4k_{[e}\tilde{\epsilon}_{bcd]}$, where the normal vector $k^a \propto \x^a$ on the horizon is future directed. The null energy condition $\d T_{ab}k^a k^b\geq 0$ ensures
\ba\label{ineq1}
\d M-\W_\math{H} \d J-\F_H\d Q\geq 0\,.
\ea
It is not difficult to know that, if \eq{ineq1} is saturated by letting $\d T_{ab}k^ak^b|_\math{H}=0$ (which means the energy flux of the first-order nonelectromagnetic perturbation vanishes), $(2M^2-Q^2)^2-4J^2\geq0$ can be violated. Accordingly, \eq{var1eq1} reduces to
\ba\label{var1eq1op}\begin{aligned}
\d M-\W_\math{H} \d J-\F_H\d Q=0\,.
\end{aligned}\ea

Next, we turn to evaluate the second-order inequality. Similarly to the analysis of the first-order perturbation, the second equation in \eq{var12} becomes
\ba\begin{aligned}
\d^2 M-\W_\math{H} \d^2 J&=-\int_\math{H}\x\.\d \bm{E}\d\f-\int_\math{H} \d^2 \bm{C}_\x+\math{E}_\S(\f,\d\f)\,.
\end{aligned}\nn\\\ea
Here the first two terms at the right-hand side only depend on $\math{H}$ as $\d \bm{E}$ and $\d^2\bm{C}_\x$ are zero on $\S_1$ in the condition that no source exists outside the black hole at very late time. Additionally,  $\x^a$ is tangent to the horizon, so the first term becomes vanishing. Considering \eq{EC}, we have
\ba
\lf(\d^2\bm{C}_\x\rt)_{abc}=\d^2\lf(\epsilon_{eabc}T_d{}^e\x^d\rt)+\frac{1}{2\sqrt{2}}\d^2\lf(\epsilon_{eabc}A_dj^e\x^d\rt)
\ea
for the second term, where we have imposed the condition $\x^a\d B_{ab}|_\math{H}=\x^a \d A_a|_\math{H}=0$ by a gauge transformation following the setting of Ref. \cite{SW}, so we have
\ba\begin{aligned}
&\frac{1}{2\sqrt{2}}\d^2\lf[\int_\math{H}\x^a A_a\epsilon_{ebcd}j^e\rt]=-\F_H \d^2\lf[\int_\math{H}\epsilon_{ebcd}j^e\rt]\\
&=-\F_\math{H}\d^2Q_\text{flux}=-\F_\math{H}\d^2Q\,,
\end{aligned}\ea
where $\d^2Q$ is the second-order change in the charge of the black hole. Furthermore, realizing that the first-order perturbation is optimal, we get
\ba\begin{aligned}\label{dM22}
\d^2M-\W_\math{H} \d^2 J&-\F_\math{H}\d^2Q=\math{E}_\S(\f,\d \f)-\int_\math{H}\x^a\epsilon_{ebcd}\d^2T_a{}^e\\
&=\math{E}_\S(\f,\d \f)+\int_\math{H}\bm{\tilde{\epsilon}}\d^2T_{ab}\x^ak^b\\
&\geq\math{E}_\math{H}(\f,\d \f)+\math{E}_{\S_1}(\f,\d \f)\, ,
\end{aligned}\ea
where the additional matter source's energy condition for the second-order perturbed stress-energy tensor have been used.

Next, we go to calculate the contribution of the horizon which can be decomposed into
\ba\label{EEE3}
\math{E}_\math{H}(\f,\d \f)=\int_\math{H}\bm{\w}^\text{GR}+\int_\math{H}\bm{\w}^\text{A}+\int_\math{H}\bm{\w}^\text{DIL}+\int_\math{H}\bm{\w}^\text{B}\,.
\ea
With similar calculations in \cite{SW} for the gravitational contribution and in \cite{jiejiang} for the electromagnetic as well as dilaton contributions, we have
\ba\begin{aligned}
&\int_\math{H}\bm{\w}^\text{GR}=\frac{1}{4\p}\int_{\math{H}}(\x^a\grad_a u)\d\s_{ac}\d\s^{bc}\bm{\tilde{\epsilon}}\geq 0\,,\\
&\int_\math{H}\bm{\w}^\text{A}+\int_\math{H}\bm{\w}^\text{DIL}=\int_\math{H}\bm{\tilde{\epsilon}}\x^ak^b\d^2\lf(T_{ab}^\text{A}+T_{ab}^\text{DIL}\rt)\geq 0\,.
\end{aligned}\ea
Then, we calculate the contribution from $\bm{B}$. From \eq{3w}, we obtain
\ba\begin{aligned}\label{wBabc}
&\bm{\w}_{abc}^\text{B}=-\frac{1}{64\p}\left\{\epsilon_{dabc}\lf[(\math{L}_\x\d\tilde{H}^{def})A_f\d A_e-\d\tilde{H}^{def}A_f\math{L}_\x\d A_e \right.\right.\\
&\left.+\tilde{H}^{def}(\math{L}_\x\d A_f)\d A_e-\tilde{H}^{def}\d A_f\math{L}_\x\d A_e-2(\math{L}_\x\d\tilde{H}^{def})\d B_{ef}\right.\\
&\left.+2\d\tilde{H}^{def}\math{L}_\x\d B_{ef}\rt]+\math{L}_\x\d\epsilon_{dabc}\lf[\tilde{H}^{def}A_f\d A_e-2\tilde{H}^{def}\d B_{ef}\right]\\
&\left.-\d\epsilon_{dabc}\lf[\tilde{H}^{def}A_f\math{L}_\x\d A_e-2\tilde{H}^{def}\math{L}_\x\d B_{ef}\right]\right\}\,.
\end{aligned}\nn\\\ea
Since the background spacetime is stationary, by considering that $H_{abc}$ is purely tangential to the horizon, the third term, fourth term, and last four terms vanish. Equation \eq{wBabc} reduces to
\ba\begin{aligned}\label{Bw}
&\bm{\w}_{abc}^\text{B}=-\frac{1}{64\p}\epsilon_{dabc}\lf[(\math{L}_\x\d\tilde{H}^{def})A_f\d A_e-2(\math{L}_\x\d\tilde{H}^{def})\d B_{ef} \right.\\
&\left.-\d\tilde{H}^{def}A_f\math{L}_\x\d A_e+2\d\tilde{H}^{def}\math{L}_\x\d B_{ef}\right]\\
&=-\frac{1}{64\p}\math{L}_\x\lf\{\epsilon_{dabc}\lf[(\d\tilde{H}^{def})A_f\d A_e-2\d\tilde{H}^{def}\d B_{ef} \right]\right\}\\
&+\frac{1}{64\p}\epsilon_{dabc}\lf[2\d\tilde{H}^{def}A_f\math{L}_\x\d A_e-4\d\tilde{H}^{def}\math{L}_\x\d B_{ef}\right]\,.
\end{aligned}\ea
Using
\ba
\math{L}_\x\bm{\h}=d(\x\.\bm{\h})
\ea
on the horizon as well as the fact that the perturbation vanishes on the bifurcation surface $B$, the first two terms on the right side contributes only a boundary term at $B_1=\math{H}\cap\S_1$. Considering the assumption that the perturbation is stationary at $B_1$, $\d \tilde{\bm{H}}$ is purely tangential to the horizon. As $\x^a \d A_a=\x^a\d B_{ab}=0$ on the horizon, the first two terms of \eq{Bw} contribute nothing to \eq{EEE3}. Then, Eq. \eq{Bw} can be reduced to
\ba\begin{aligned}\label{Bw2}
\bm{\w}_{abc}^\text{B}=\frac{1}{32\p}\epsilon_{dabc}\lf[\d\tilde{H}^{def}A_f\d F_{ge}\x^g-2\d\tilde{H}^{def}\d (d \bm{B})_{gef}\x^g\right]\,,\\
\end{aligned}\nn\\\ea

Meanwhile, according to \eq{TTT}, we have
\ba\begin{aligned}
&\epsilon^d{}_{abc}\x^e\d^2 T^\text{B}_{de}=\frac{1}{16\p}\epsilon_{dabc}\d\tilde{H}^{def}\d H_{gef} \x^g\\
&=\frac{1}{16\p}\epsilon_{dabc}\d\tilde{H}^{def}\lf[\d (d\bm{B})_{gef} \x^g-\frac{1}{2} A_e\d F_{fg}\x^g\right.\\
&\left.-\frac{1}{2}\d A_e F_{fg}\x^g-\frac{1}{4}A_g\x^g\d F_{ef}\right]\\
&=\frac{1}{16\p}\epsilon_{dabc}\d\tilde{H}^{def}\lf[\d (d\bm{B})_{gef} \x^g-\frac{1}{2} A_e\d F_{fg}\x^g\right]\\
&-\frac{1}{32\p}\epsilon_{dabc}\lf[2\d\tilde{H}^{def}\d A_e F_{fg}-\grad_e\lf(\d\tilde{H}^{def}A_g\x^g\d A_f\rt)\x^g\rt]\,.
\end{aligned}\ea
Here we have used the fact $\math{L}_\x \bm{A}=0$ for the background field and the collision matter does not have the source of the gauge field $\bm{B}$, i.e.,
\ba
\grad_a\d \tilde{H}^{abc}=0\,.
\ea
Since the last term can be expressed as the boundary term of the horizon, as well as the stationary conditions, the last two terms vanish. Therefore, it is not hard to see
\ba
\int_\math{H}\bm{\w}^\text{B}=\int_\math{H}\bm{\tilde{\epsilon}}\x^ak^b\d^2T^\text{B}_{ab}\,.
\ea
Also we have
\ba
\math{E}_\math{H}(\f,\d\f)=\int_\math{H}\bm{\tilde{\epsilon}}\x^ak^b\d^2\lf(T_{ab}^\text{A}+T_{ab}^\text{B}+T_{ab}^\text{DIL}\rt)\geq 0\,,
\ea
where the  null energy condition of the gauge fields' and dilaton field's stress-energy tensors have been used. With these in mind, \eq{dM22} reduces to
\ba\label{ineq2xx}
\d^2M-\W_\math{H}\d^2J-\F_H\d^2Q\geq \math{E}_{\S_1}(\f,\d\f)\,.
\ea
Now we are going to compute $\math{E}_{\S_1}(\f,\d\f)$. Following the procedures used in \cite{SW}, we can evaluate it through $\math{E}_{\S_1}(\f,\d\f)=\math{E}_{\S_1}(\f,\d\f^\text{KS})$, with $\f^\text{KS}$ being introduced by the variations of Kerr-Sen black hole solutions \eq{ds2},
\ba\label{varMQ}\begin{aligned}
M^\text{KS}(\l)&=M+\l \d M\,,\\
J^\text{KS}(\l)&=J+\l \d J\,,\\
Q^\text{KS}(\l)&=Q+\l \d Q\,,
\end{aligned}\ea
in which $\d M$, $\d Q$ and $\d J$ are at the same order with the above optimal perturbation of the matter source, i.e., they are both at the first order. One can get $\d^2 M=\d^2Q=\d^2 J=\d \bm{E}_\f=\d^2 \bm{C}=\math{E}_\math{H}(\f,\d \f^\text{KS})=0$ from \eq{varMQ}. Therefore, according to \eq{var12}, we can obtain
\ba
\math{E}_{\S_1}(\f, \d \f^\text{KS})=-\int_B\lf[\d^2\bm{Q}_\x-\x\.\d \bm{\Q}(\f,\f^\text{KS})\rt]\,.
\ea
As $\x^a=0$ on $B$, the above equation can be written as
\ba
\math{E}_{\S_1}(\f, \d \f^\text{KS})=-\frac{\k}{8\p}\d^2A_B^\text{KS}\,.
\ea
Thus the second-order inequality \eq{ineq2xx} reduces to
\ba\label{ineq111}
\d^2M-\W_\math{H}\d^2J-\F_\math{H}\d^2 Q\geq -\frac{\k}{8\p}\d^2A_B^\text{KS}\,.
\ea
Taking variations of Eq. \eq{kAWF} twice, we can evaluate the r.h.s of this inequality. Together with the optimal first-order inequality $\d M=\W_\math{H}\d J+\F_\math{H}\d Q$, $\d^2A_B^\text{KS}$ can be given by
\ba\begin{aligned}
&\d^2A_B^\text{KS}=-\frac{\p}{M^4\epsilon} \lf[(2Q^2+4M^2+4\epsilon M^2)\d J^2\right.\\
&\left.-8J Q \d J\d Q+\lf(2Q^4+8M^4(1+\epsilon)-4M^2Q^2(2+\epsilon)\rt)\rt]\,,
\end{aligned}\ea
in which
\ba
\epsilon=\sqrt{\lf(1-\frac{Q^2}{2M^2}\rt)^2-\frac{J^2}{M^3}}\,.
\ea
For the near-extremal black hole, $\epsilon$ is a small parameter. Then, the surface gravity of this black hole can be given by
\ba
\k=\frac{M\epsilon}{2M^2(1+\epsilon)-Q^2}\,.
\ea

Expanding the right-hand side of the second perturbation inequality \eq{ineq111} to the lowest order in $\epsilon$, we have
\ba\label{ineq2}\begin{aligned}
&\d^2M-\W_\math{H}\d^2J-\F_\math{H}\d^2 Q\geq \frac{1}{4M^3(2M^2-Q^2)}\\
&\times\lf[(2M^2+Q^2)\d J^2-4Q J \d J\d Q+(2M^2-Q^2)^2\d Q^2\rt]\,.
\end{aligned}\ea

\section{Gedanken experiments to destroy a nearly extremal Kerr-Sen black hole}\label{sec5}
Now we shall investigate the gedanken experiments overcharging a near-extremal Kerr-Sen black hole by the physical process introduced above. A function of $\l$ can be defined as
\ba
h(\l)=\lf[2M(\l)^2-Q(\l)^2\rt]^2-4J(\l)^2\,.
\ea
The WCCC is violated if there exists a spacetime solution $\f(\l)$ such that $h(\l)<0$. Considering the second-order approximation of $\l$, we have
\ba\begin{aligned}
&h(\l)=(2M^2-Q^2)^2-4J^2\\
&+4\lf[(2M^2-Q^2)(2M \d M-Q\d Q)-2J\d J\rt]\l\\
&+2\lf[(2M^2-Q^2)(2M \d^2 M-Q\d^2 Q)-2J\d^2 J\rt]\l^2\\
&+\lf[4(6M^2-Q^2)\d M^2-2(2M^2-3Q^2)\d Q^2\right.\\
&\left.-4 \d J^2-16 M Q \d M\d Q\rt]\l^2\,.
\end{aligned}\ea

In this paper, we would like to consider the case when the background spacetime is near extremal, and the small parameter $\epsilon$ is chosen to be in agreement with the first-order perturbation of the matter source.

First, we would like to analyze the result found in \cite{Yu:2018eqq} for the old version of gedanken experiments. In that scene, we would like to neglect the $O(\l^2)$ term. Then, using the first-order perturbation inequality \eq{ineq1}, we have
\ba
f(\l)\geq 4M^2\epsilon \lf(M^2\epsilon-2\d J\rt)\,,
\ea
which implies that it is possible to make $f(\l)<0$. Therefore, if we neglect the second-order perturbation, the Kerr-Sen black hole can be overspun or overcharged.

Next, let us turn to the gedanken experiment in its new version. As long as \eq{var1eq1op} and \eq{ineq2} are used, on the level of second-order approximation, we immediately achieve
\ba
h(\l)\geq 4(\d J-M^2 \epsilon)^2>0\,.
\ea
Thus, as we can see, when the second-order correction of the perturbation is taken into account, this Kerr-Sen black hole cannot be overspun or overcharged. There is no violation of WCCC around the Kerr-Sen black holes.

\section{Conclusion}\label{sec6}

In this paper, we have adopted the new version of the gedanken experiments proposed by Source and Wald in \cite{SW} to test the weak cosmic censorship conjecture for Kerr-Sen black holes. First of all, we derive the first two order perturbation inequalities in the low energy limit of heterotic string theory based on the Iyer-Wald formalism, and apply it into the Kerr-Sen black holes. Differing from the result of the old version of gedanken experiments where the Kerr-Sen black holes can be destroyed if neglecting the self-force, self-energy, and radiative effect \cite{Siahaan,Gwak:2016gwj}, it cannot be overcharged or overspun after the second-order perturbation inequality being considered.  Therefore, WCCC is well preserved around this black hole. This result might indicate that an already existing Kerr-Sen black hole will never be destroyed classically. Moreover, the above results support the suggestion that the second-order perturbation inequality acts as the self-energy or backreaction for the collision matter.

\section*{acknowledgements}
This research was supported by National Natural Science Foundation of China (NSFC) with Grant No. 11675015.

\end{document}